\documentclass[a4paper,pra,reprint, twocolumn,superscriptaddress]{revtex4-2}
\usepackage{ulem}
\usepackage{amssymb}
\usepackage{amsmath}
\usepackage{epsfig}
\usepackage{color}
\usepackage{graphics, graphicx}
\usepackage{bbold}
\usepackage{psfrag}
\usepackage{mathcomp}
\usepackage{amsmath}
\usepackage{amssymb}%
\usepackage{mathrsfs}%
\usepackage{subfigure}
\usepackage{verbatim}
\usepackage{xcolor}
\usepackage[colorlinks,citecolor=blue,urlcolor=blue]{hyperref}

\usepackage{mathrsfs}

\begin{document}

\date{\today}
\title{Molecular state in a spin-orbital-angular-momentum coupled Fermi gas}
\author{Yiwen Han}
\affiliation{CAS Key Laboratory of Quantum Information, University of Science and Technology of China, Hefei 230026, China}
\author{Shi-Guo Peng}
\affiliation{State Key Laboratory of Magnetic Resonance and Atomic and Molecular Physics, Innovation Academy for Precision Measurement Science and Technology, Chinese Academy of Sciences, Wuhan 430071, China}
\author{Ke-Ji Chen}
\affiliation{Key Laboratory of Optical Field Manipulation of Zhejiang Province and Physics Department of Zhejiang Sci-Tech University, Hangzhou 310018, China}
\author{Wei Yi}
\email{wyiz@ustc.edu.cn}
\affiliation{CAS Key Laboratory of Quantum Information, University of Science and Technology of China, Hefei 230026, China}
\affiliation{CAS Center For Excellence in Quantum Information and Quantum Physics, Hefei 230026, China}

\begin{abstract}
We study the two-body bound states in a spin-orbital-angular-momentum (SOAM) coupled quantum gas of fermions.
Two different configurations are considered: an attractive $s$-wave interaction exists between two spin species that are SOAM coupled; and an atom with SOAM coupled internal spins interacts state-selectively with another atom. For both cases, we identify the condition for the emergence of molecular states with finite total angular momenta.
These molecular states with quantized  total angular momenta
correspond to the SOAM-coupling-induced vortices in the corresponding Fermi superfluid.
We propose to detect the molecules through Raman spectroscopy with Laguerre-Gaussian lasers.
As the molecular states can form above the superfluid temperature, they offer an experimentally more accessible route toward the study of the underlying pairing mechanism under SOAM coupling.
\end{abstract}

\maketitle

\section{Introduction}

Synthetic spin-orbit coupling in cold atoms modifies the single-particle dispersion, and offers a powerful tool for quantum control~\cite{Lin-11,Zhang-12,Zwierlein-12}. In the past decade, such control has enabled the simulation of band topology, and led to the prediction of various exotic few- and many-body states~\cite{socreview1,socreview2,socreview3,socreview4,socreview5,socreview6}.
Building upon these achievements, the recent experimental realization of spin-orbital-angular-momentum coupling (SOAMC) introduces even more opportunities~\cite{Jiang-19,Lin-18,Hu-15,Pu-15,Sun-15,Qu-15,Chen-16,Hu-19,Chen-19,Han-20,Duan-20}. Therein, different ground hyperfine states of an atom are coupled by a pair of co-propagating Laguerre-Gaussian beams with distinct orbital angular momentum,
giving rise to the experimental observation of spin-dependent vortices in spinor Bose-Einstein condensates under SOAMC~\cite{Jiang-19,Lin-18}.
Further, as a direct consequence of the deformed single-particle dispersion in the discretized angular-momentum space, a unique vortex-forming mechanism exists in the SOAM coupled Fermi superfluids~\cite{Chen-20,Wang-21}. And it has been proposed very recently that an angular topological superfluid can be induced by SOAMC~\cite{Chen-22}, whose topological defect, in the form of giant vortices, has interesting implications for topological quantum computation.

Nevertheless, a fundamental hurdle to the experimental observation of the SOAMC-induced pairing states is the inevitable heating introduced by the Raman process~\cite{heating}, which makes it difficult to cool the system below the superfluid temperature. It is exactly due to this reason that, despite a plethora of theoretical study on exotic pairing states in spin-orbit coupled Fermi gases, none have so far been observed in experiments. Instead, it is much easier to generate dressed molecules under spin-orbit coupling, which persist above the critical temperature~\cite{spielmandressed,zhangdressed}. Similarly, it is reasonable to expect that molecular states in an SOAM coupled Fermi gas should be readily accessible under typical experimental conditions, and would provide much desired insight into the pairing mechanism under SOAMC.

In this work, we study in detail the molecular state in an SOAM coupled Fermi gas, and show that two-body bound states with a quantized total angular momentum can be stabilized, leaving detectable signatures in the Raman spectroscopy. We consider two different scenarios. In the first case, a molecular state is formed through an attractive $s$-wave interaction between two hyperfine spin states that are also coupled through the SOAMC. The molecule acquires a finite total angular momentum under a sufficiently strong Zeeman field that breaks the time-reversal symmetry. The underlying pairing mechanism is similar to that of the SOAMC-induced vortex state in Ref.~\cite{Chen-20,Wang-21}, and is the angular analogue of the spin-orbit-coupling-induced Fulde-Ferrell state in Ref.~\cite{Wu-13}.
In the second case, SOAMC is only enforced upon the internal states of one of the atoms, which interacts with the other atom in a spin-selective fashion. The resulting molecular state also acquires a finite angular momentum, due to the interplay between SOAMC and spin-selective interaction~\cite{ZCY-14}.
We then propose to detect the molecular state, particularly their quantized angular momentum, through Raman spectroscopy with Laguerre-Gaussian beams.
Our results suggest that understanding molecules is a natural first step in the experimental study of the unique pairing states under SOAMC.

The paper is organized as follows. In Sec.~II, we present the models for the two scenarios. We focus on the characterization of the first scenario in Sec.~III, emphasizing the quantized total angular momentum of the molecular state and its signal in the direct Raman spectroscopy. In Sec.~IV, we focus on the characterization of the second scenario, where we propose a detection scheme based on inverse Raman spectroscopy. We summarize in Sec.~V.


\begin{figure}[tbp]
\includegraphics[width=0.48\textwidth]{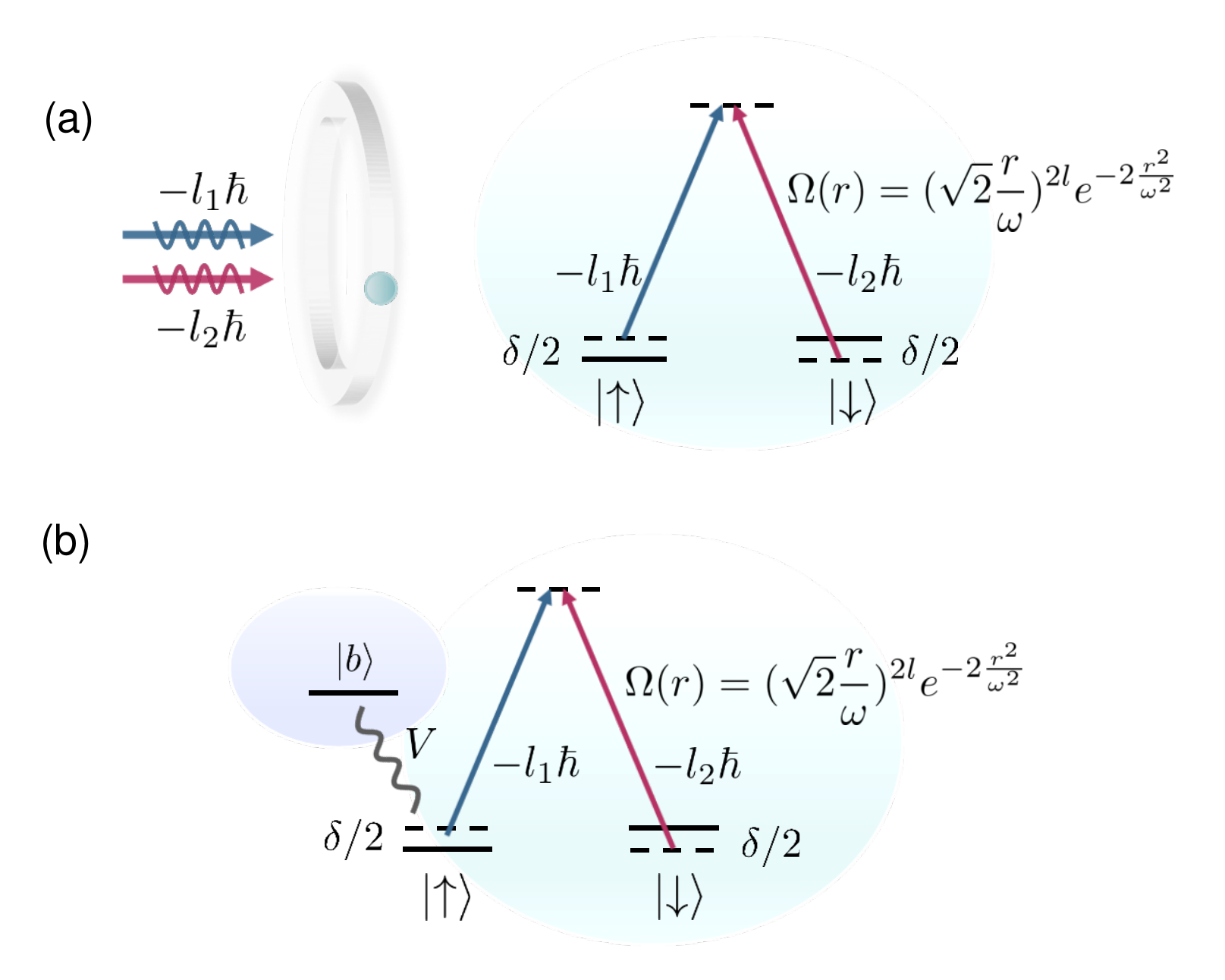}
\caption{(a) Schematic illustration of the first scenario, where two spin species with attractive $s$-wave interaction are Raman coupled by a pair of co-propagating Laguerre-Gaussian beams. Due to the spatial light distribution of the lasers, the atoms are effectively trapped in a ring geometry under the a.c. Stark potential.
(b) Schematic illustration of the second scenario, where the same pair of Laguerre-Gaussian beams couple the internal spin states $\left|\uparrow\right\rangle$ and $\left|\downarrow\right\rangle$ of the first atom. The other atom (labeled $|b\rangle$) interacts only with the spin-up species.
See discussions in the main text for definitions of the various variables.
}
\label{fig:fig1}
\end{figure}

\section{Model}

As illustrated in Fig.~\ref{fig:fig1}, we consider two different scenarios. In the first scenario [Fig.~\ref{fig:fig1}(a)], each of the two atoms is subject to the same SOAMC, generated by a pair of Laguerre-Gaussian beams with different orbital angular momentum. While the atoms are tightly confined in the $x$--$y$ plane (in a quasi-two-dimensional potential), the spatial profiles of the overall light field induces a ring-shape a.c. Stark potential within the plane, which restricts the radial atomic motion. An attractive $s$-wave interaction exists between the two spin species.

In the second scenario [Fig.~\ref{fig:fig1}(b)], SOAMC is only applied on the internal spin states of one of the atoms, whereas the other atom (in state $\left|b\right\rangle$) interacts only with the spin-up species of the first atom. Thus, the second atom acts as a probe through which information of the SOAMC is reflected in the resulting molecular state.

\begin{figure}[tbp]
\includegraphics[width=0.48\textwidth]{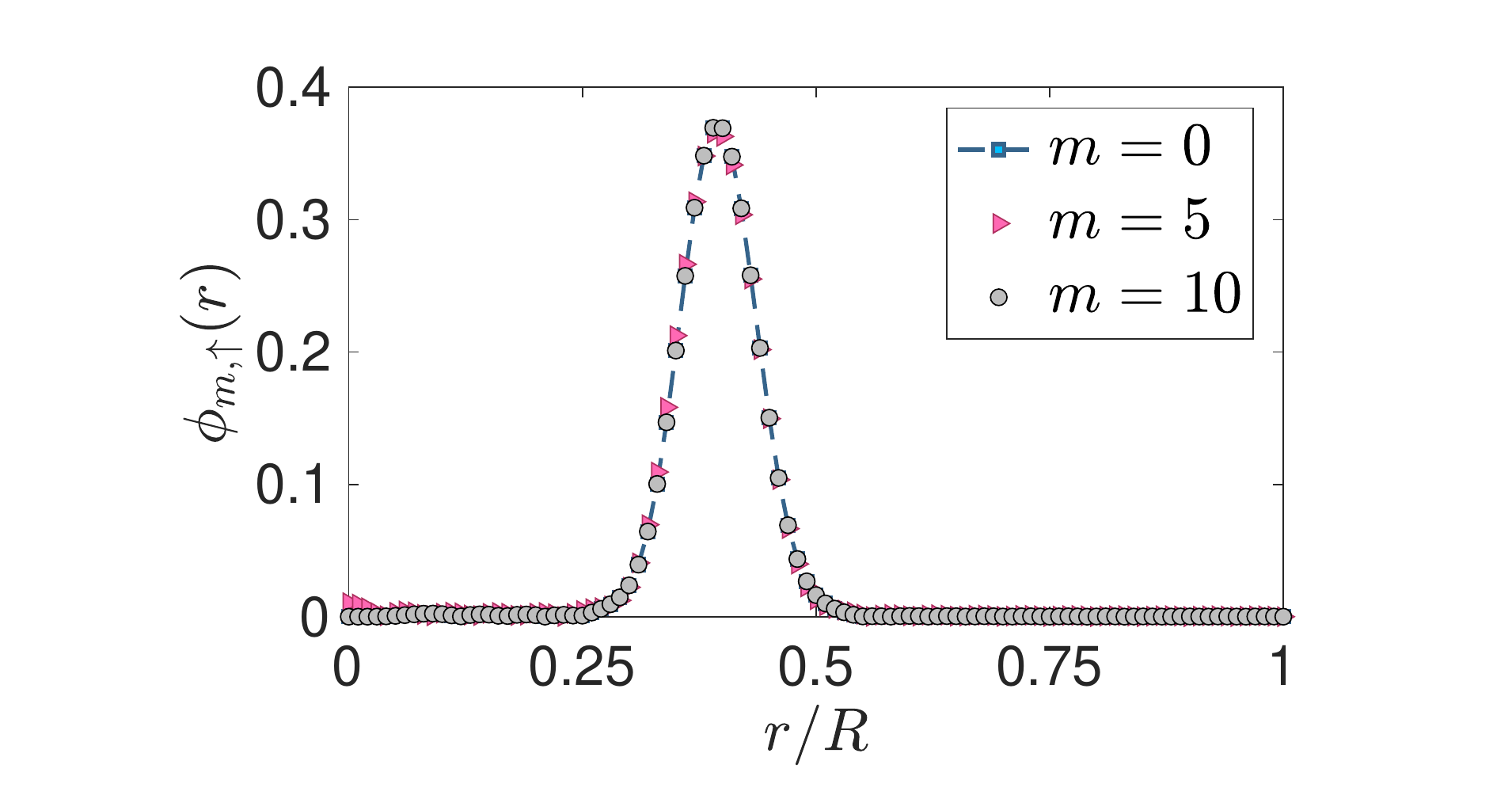}
\caption{Radial distribution of the eigen wavefunction $\phi_{m,\uparrow}(r)$ for different angular-momentum modes. For numerical calculations, we take $l=5$, $\delta=0$, $R/w=4$, and $\chi/E_l=-1$. Here $E_l$ is the unit of energy defined in the main text.
}
\label{fig:fig2}
\end{figure}

\subsection{Hamiltonian for scenario I}

We first write down the single-particle Hamiltonian for the first scenario in the spin basis $\{\left|\uparrow\right\rangle,\left|\downarrow\right\rangle\}$
\begin{align}
H_{0}=&-\frac{\hbar^{2}\nabla^{2}}{2M}+V_{\text{ext}}(\mathbf{r})+\frac{(l\hbar)^{2}}{2M r^{2}}+\chi I(r)+\Omega I(r)\sigma_{x}\nonumber\\
&-\frac{l\hbar}{M r^{2}}L_{z}\sigma_{z}+\frac{\delta}{2}\sigma_{z},\label{eq:H0}
\end{align}
where $\sigma_{i}$ ($i=x,y,z$) are the Pauli matrices, $M$ is the atomic mass, $L_{z}=-i\hbar \partial /\partial \theta$ is the atomic angular momentum operator in the polar coordinate ${\bf r}=(r,\theta)$, $2l\hbar=(l_1-l_2)\hbar$ is the transferred orbital angular momentum, and $\delta$ is the two-photon detuning of the Raman process generating the SOAMC. To derive the Hamiltonian, we have taken a gauge transformation $U={\rm diag}(e^{-il \theta},\  e^{il \theta})$, and assumed a hard-wall box potential $V_{\text{ext}}({\bf r})$ with a radius $R$.
Under the spatially inhomogeneous Laguerre-Gaussian beams, the atoms are subject to the Raman and a.c. Stark potentials, denoted as $\Omega I(r)$ and $\chi I(r)$, respectively, with $I(r)=\left(\sqrt{2}r/\omega\right)^{2l}e^{-2r^{2}/\omega^{2}}$. Here $w$ is the laser waist.

Throughout this work, we assume that the a.c. Stark potential is sufficiently deep, such that the atomic radial degrees of freedom are frozen into the ground state with $n=1$. Key features of the molecular state (or pairing states in the many-body case) are preserved in this single-mode approximation~\cite{Chen-22}.

It is then convenient to adopt a second-quantized form of Hamiltonian (\ref{eq:H0}), by expanding the field operators of the two spin species as $\psi_{\sigma}(\mathbf{r})=\sum_m \phi_{m,\sigma}(r)\Theta(\theta)a_{m,\sigma}$. Here $a_{m,\sigma}$ are the annihilation operators of the corresponding mode, and $\{\phi_{m\sigma}\}$ are the eigen wavefunctions of  $K_{\sigma}= -\frac{\hbar^2}{2M} \left[\frac{1}{r}\frac{\partial}{\partial r}(r\frac{\partial}{\partial r})+\frac{1}{r^2}\left(\frac{\partial}{\partial \theta}-i\tau l \right)^2 \right]+\chi I(r)+\tau\frac{\delta}{2}$, with $\tau=+1\ (-1)$ for $\sigma=\uparrow(\downarrow)$. Specifically, we have
\begin{align}
K_{\sigma}(\mathbf{r})\phi_{m,\sigma}(r)\Theta_{m}(\theta)=\epsilon_{m,\sigma}\phi_{m,\sigma}(r)\Theta_{m}(\theta),
\end{align}
where $\epsilon_{m,\sigma}$ is the corresponding eigenvalue.

The second-quantized Hamiltonian is then
\begin{align}
H_{s}=\sum_{m\sigma}\epsilon_{m,\sigma}a_{m,\sigma}^{\dagger}a_{m,\sigma}+\sum_m \Omega_{m}\left(a_{m,\uparrow}^{\dagger}a_{m,\downarrow}+a_{m,\downarrow}^{\dagger}a_{m,\uparrow}\right),
\end{align}
where
\begin{align}
\Omega_{m}=\Omega\int rdr\phi_{m,\uparrow}(r)I(r)\phi_{m,\downarrow}(r).
\end{align}

\begin{figure*}[tbp]
\includegraphics[width=0.8\textwidth]{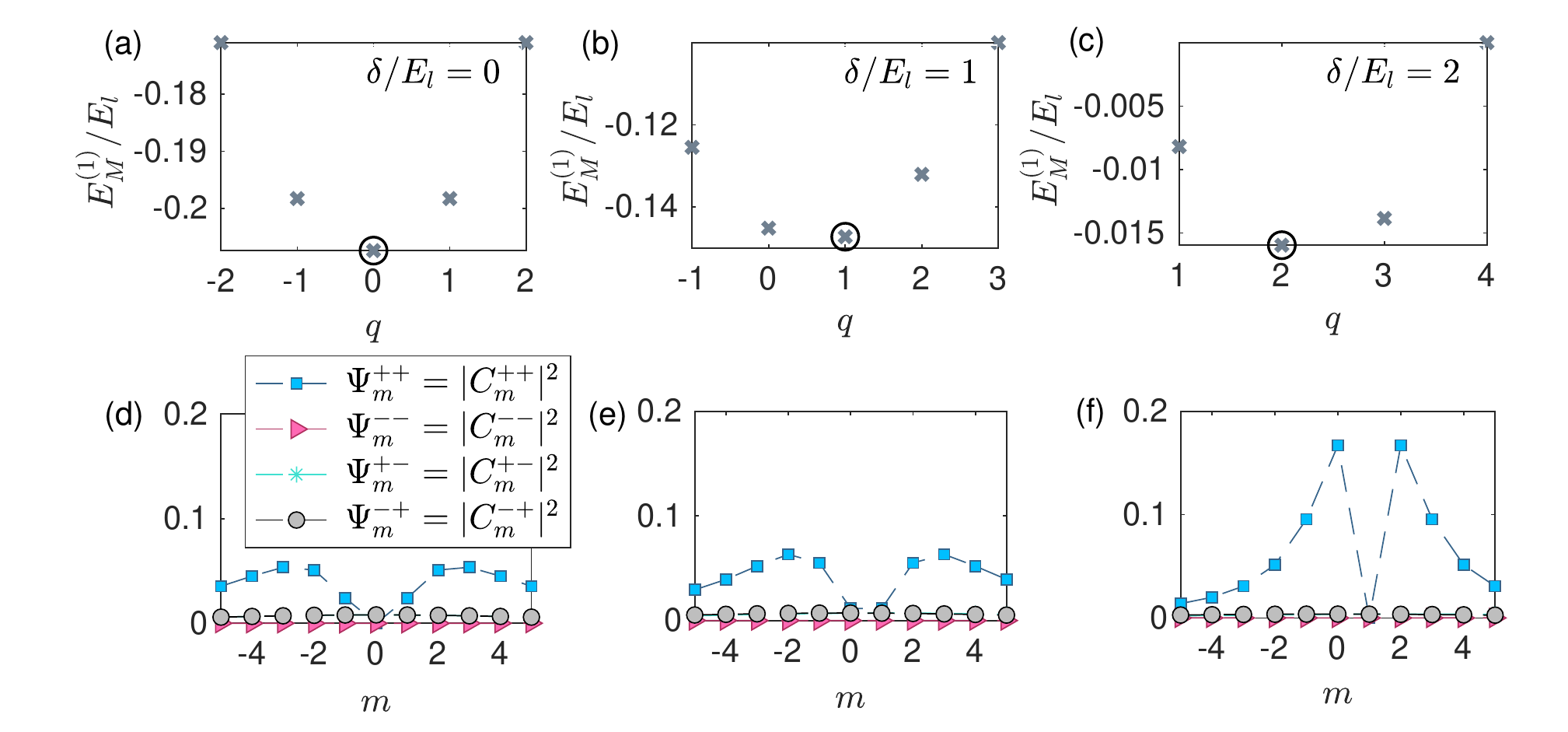}
\caption{(a)(b)(c) The molecular binding energy $E_M^{(1)}$ as functions of the total angular momentum $q$, under different detunings: (a) $\delta=0$, (b) $\delta/E_l=1$, (c) $\delta/E_l=2$.
The ground states feature (a) $q=0$, (b) $q=1$, (c) $q=2$.
(d)(e)(f) The ground-state molecular wavefunction in the discretized angular-momentum space (characterized by $m$), under the parameters in (a)(b)(c), respectively. For all calculations, we fix $l=5$, $E_b^{(1)}/E_l=-6$, $\Omega/E_l=0.18$.
}
\label{fig:fig3}
\end{figure*}

\begin{figure}[tbp]
\includegraphics[width=0.48\textwidth]{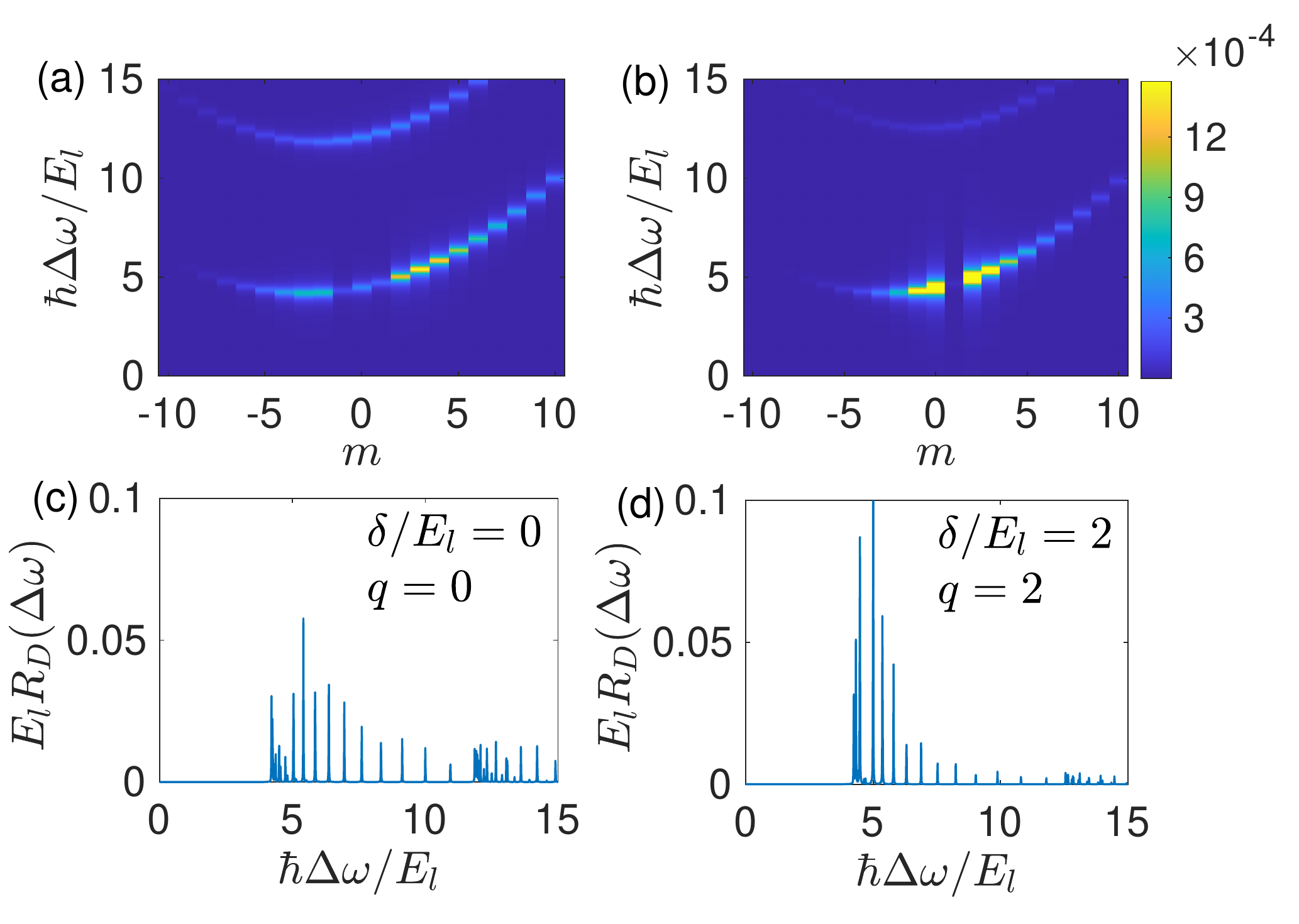}
\caption{Direct Raman spectrum. (a) Angular-momentum resolved spectrum $R_\text{D}(m,\Delta\omega)$ for a molecular state with $q=0$.
(b) Angular-momentum resolved spectrum for a molecular state with $q=2$.
(c) Integrated spectrum $R_\text{D}(\Delta\omega)$ for the state in (a).
(d) Integrated spectrum for the state in (b).
Other parameters are $l=5$, $E_b^{(1)}/E_l=-6$, and $\Omega/E_{l}=0.18$.
}
\label{fig:fig4}
\end{figure}

As illustrated in Fig.~\ref{fig:fig2}, when the a.c. Stark potential is sufficiently deep, the radial mode functions $\phi_{m\sigma}$ for different angular-momentum modes $m$ overlap with each other, and are all concentrated on a ring, with radius $r_0=\sqrt{l\omega^2/2}$. The Hamiltonian can then be further simplified as
\begin{align}
H_{\text{ring}}=\sum_{m\sigma}\xi_{m,\sigma}a_{m,\sigma}^{\dagger}a_{m,\sigma}+h\sum_m \left(a_{m,\uparrow}^{\dagger}a_{m,\downarrow}+a_{m,\downarrow}^{\dagger}a_{m,\uparrow}\right),\label{eq:Hsingle}
\end{align}
where $\xi_{m,\uparrow(\downarrow)}=\xi_{m}\pm(\alpha m+\frac{\delta}{2})$, $\xi_{m}=m^{2}\hbar^{2}/(2Mr_{0}^{2})$, $\alpha=l\hbar^{2}/(Mr_{0}^{2})$, and $h=\Omega I(r_{0})$. Throughout the work, we will use $E_l=\alpha l$ as the unit of energy.

Hamiltonian (\ref{eq:Hsingle}) is easily diagonalized, with $H_{\text{ring}}=\sum_{m,\beta=\pm}\lambda_{m,\beta}a^\dag_{m,\beta}a_{m,\beta}$. Here $a_{m,\beta}$ are the annihilation operators of the SOAM coupled helicity bands, with $\lambda_{m,\pm}=\xi_{m}\mp\sqrt{h^{2}+(\alpha m+\delta/2)^{2}}$. These expressions are the angular analogue of those in the momentum space of a one-dimensional system under spin-orbit coupling.

We consider a contact $s$-wave interaction between the two spin species. The interaction term under the single-mode approximation is written as
\begin{equation}
H_{\text{int}}=V \sum_{mm'q}a_{m',\uparrow}^{\dagger}a_{q-m',\downarrow}^{\dagger}a_{q-m,\downarrow}a_{m,\uparrow},
\end{equation}
where we have dropped the angular dependence of $V$, since $\phi_{m,\sigma}(r)$ almost overlap for different angular-momentum modes $m$. We renormalize the bare interaction rate $V$ through~\cite{Chen-20,Chen-22}
\begin{align}
\frac{1}{V}=\sum_m\frac{1}{E_b^{(1)}-\epsilon_{m,\uparrow}-\epsilon_{-m,\downarrow}},
\end{align}
where $E_b^{(1)}$ is the energy of the two-body bound state in the same ring geometry but without SOAMC.

\subsection{Hamiltonian for scenario II}

Similarly, under the single-mode approximation, the Hamiltonian for scenario II is written as
\begin{align}
H^{(2)}=&\sum_{m\sigma}\xi_{m,\sigma}a_{m,\sigma}^{\dagger}a_{m,\sigma}+h\sum_m\left(a_{m,\uparrow}^{\dagger}a_{m,\downarrow}+a_{m,\downarrow}^{\dagger}a_{m,\uparrow}\right)\nonumber\\
 & +\sum_{m}\epsilon_{m}^{b}b_{m}^{\dagger}b_{m}+g\sum_{mm'q}a_{m,\uparrow}^{\dagger}b_{q-m}^{\dagger}b_{q-m'}a_{m',\uparrow},
\end{align}
where $\epsilon_{m}^{b}=m^{2}\hbar^{2}/(2Mr_{0}^{2})$. Here we have assumed that the two atoms have the same mass $M$, and taken the two-photon detuning $\delta=0$.
The bare interaction rate $g$ is renormalized through
\begin{align}
\frac{1}{g}=\sum_m\frac{1}{E_b^{(2)}-\epsilon_{m,\uparrow}-\epsilon_{q-m}^{b}},
\end{align}
where $E_b^{(2)}$ is the two-body bound state energy under the same setup but without SOAMC.

\section{Molecular state for scenario I}

We are now in a position to solve for the molecular state. For the first scenario, the molecular state can be written as
\begin{align}
|\Psi_{q}^{(1)}\rangle=\displaystyle{\sum_m\sum_{\alpha,\beta=\pm}}C_{m}^{\alpha\beta}a_{m,\alpha}^{\dagger}a_{q-m,\beta}^{\dagger}\left|0\right\rangle,
\end{align}
where $C_m^{\alpha\beta}$ is the wavefunction in the angular-momentum space, $q$ is the total angular momentum, and $|0\rangle$ is the vacuum sate. From the Schr\"odinger's equation $(H_{\text{ring}}+H_{\text{int}})|\Psi_q^{(1)}\rangle=E_q^{(1)}|\Psi_q^{(1)}\rangle$, we derive the closed equation
\begin{align}
\frac{2}{V}= &\sum_{m}\big[\frac{\left(u_{m}v_{q-m}-v_{m}u_{q-m}\right)^{2}}{E_q^{(1)}-\lambda_{m,+}-\lambda_{q-m,+}}+\frac{\left(u_{m}v_{q-m}-v_{m}u_{q-m}\right)^{2}}{E_q^{(1)}-\lambda_{m,-}-\lambda_{q-m,-}}\nonumber\\
 & +\frac{\left(u_{m}u_{q-m}+v_{m}v_{q-m}\right)^{2}}{E_q^{(1)}-\lambda_{m,+}-\lambda_{q-m,-}}+\frac{\left(u_{m}u_{q-m}+v_{m}v_{q-m}\right)^{2}}{E_q^{(1)}-\lambda_{q-m,+}-\lambda_{m,-}}\big].
\end{align}
Here the coefficients $\{u_m,v_m\}$ relate the spin- and helicity basis, and are given by
\begin{align}
u_{m}&=\sqrt{\frac{1}{2}\Big[1-\frac{\alpha m+\delta/2}{\sqrt{h^{2}+(\alpha m+\delta/2)^{2}}}\Big]},\label{eq:u}\\
v_{m}&=\sqrt{\frac{1}{2}\Big[1+\frac{\alpha m+\delta/2}{\sqrt{h^{2}+(\alpha m+\delta/2)^{2}}}\Big]}.\label{eq:v}
\end{align}
The molecular binding energy is then $E_M^{(1)}=E_q^{(1)}-E_{\text{th}}$, where the threshold energy of the two-body scattering states is $E_{\text{th}}=2\min (\lambda_{m,+})$.

In Fig.~\ref{fig:fig3}(a)(b)(c), we show the calculated binding energy $E_M^{(1)}$ as a function of the angular momentum $q$, under different detunings. While the distribution is generally asymmetric with respect to $q=0$ under a finite detuning [Fig.~\ref{fig:fig3}(b)(c)], the ground state lies in a finite $q$ sector under a sufficiently large detuning [Fig.~\ref{fig:fig3}(c)]. This picture is consistent with the vortex state in an SOAM coupled Fermi superfluid, which starts to emerge above a critical detuning. In Fig.~\ref{fig:fig3}(d)(e)(f), we further plot the angular-momentum distribution of ground-state wavefunctions, where the location of the dip in $|C_m^{++}|^2$ corresponds to one half the quantized total angular momentum of the molecule. The dip is directly related to the induced $p$-wave pairing symmetry under SOAMC.

We propose to detect the molecular state through a direct Raman spectroscopy, in which a two-photon Raman process couples the spin-down state to a third state $|3\rangle$ that is not interacting with either spin species. For a coupling term $H_\text{R}=\Omega_\text{R}\sum_{m}e^{i\Delta\omega t}a^\dag_{m,3}e^{il \theta}a_{m,\downarrow}$, the angular-momentum resolved rate of transition is given by~\cite{Peng-12}
\begin{align}
R_{\text{D}}(m,\Delta\omega)=&-\displaystyle{\sum_{m',\beta=\pm}}\text{Im}\frac{\left|\left\langle m',\beta\left|a_{m\downarrow}\right|\Psi_{q}^{(1)}\right\rangle \right|^{2}}{\Delta\omega-(\lambda_{m',\beta}+\xi_{m+l}-E_{q}^{(1)})+i0^+},\nonumber\\
=&4\pi\underset{\beta=\pm}{\sum}|C_{m}^{+\beta}\nu_{m}-C_{m}^{-\beta}u_{m}|^{2}\nonumber\\
&\times\delta\left[\Delta\omega-(\lambda_{q-m,\beta}+\xi_{m+l}-E_{q}^{(1)})\right].
\end{align}
Here $|m,\beta\rangle=a_{m,\beta}^\dag|0\rangle$, $\Omega_\text{R}$ is the amplitude of the coupling term, the factor $e^{il\theta}$ comes from the gauge transformation, and $\Delta\omega$ is the detuning of two-photon coupling under an appropriate rotating frame where $\left|\downarrow\right\rangle$ and $|3\rangle$ are degenerate in the absence of SOAMC and interaction.
The overall transition rate is $R_\text{D}(\Delta\omega)=\sum_m R_\text{D}(m,\Delta\omega)$.

\begin{figure*}[tbp]
\includegraphics[width=0.8\textwidth]{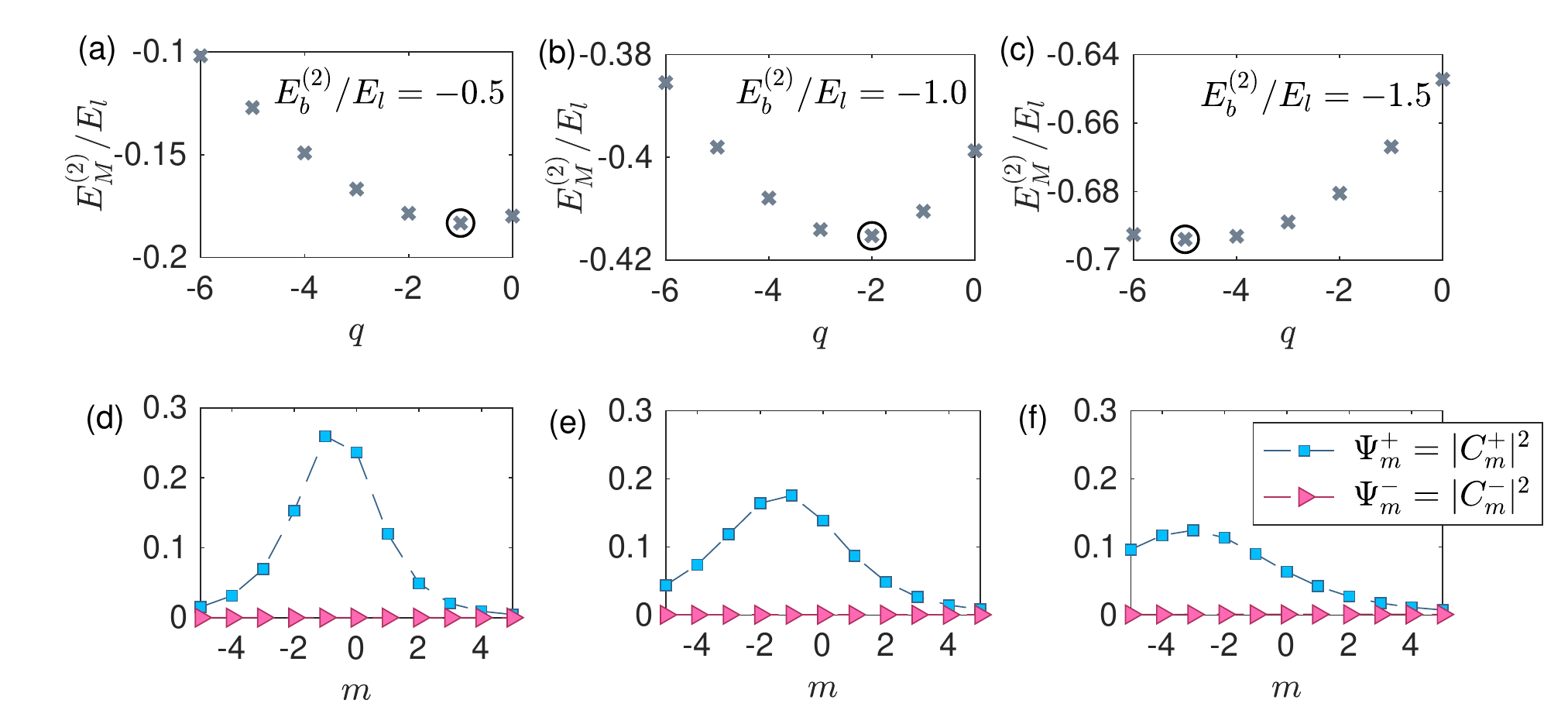}
\caption{(a)(b)(c) The molecular binding energy $E_M^{(2)}$ as functions of the total angular momentum $q$, under interaction strenghts: (a) $E_b^{(2)}/E_l=-0.5$, (b) $E_b^{(2)}/E_l=-1.0$, (c) $E_b^{(2)}/E_l=-1.5$.
(d)(e)(f) The ground-state molecular wavefunction in the discretized angular-momentum space (characterized by $m$), under the parameters in (a)(b)(c), respectively. For all calculations, we fix $l=5$, and $\Omega/E_l=0.2$.
}
\label{fig:fig5}
\end{figure*}

\begin{figure}[tbp]
\includegraphics[width=0.48\textwidth]{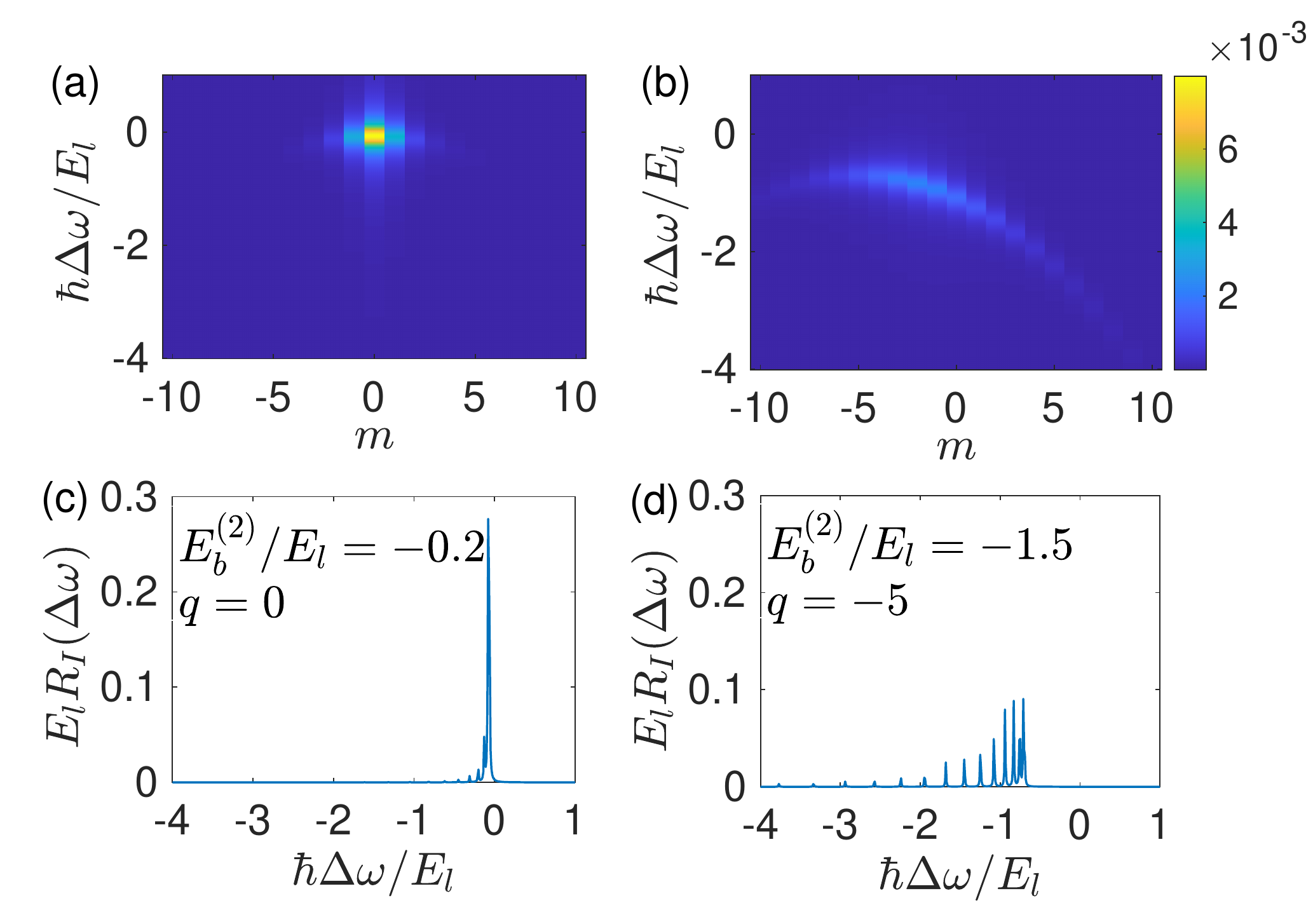}
\caption{(a) Inverse angular-momentum resolved Raman spectrum for a molecular state with $q=0$ and $E_b^{(2)}/E_l=-0.2$.
(b) Inverse angular-momentum resolved Raman spectrum for a molecular state with $q=-5$ and $E_b^{(2)}/E_l=-1.5$.
(c) and (d) are the integrated inverse spectra $R_\text{I}(\Delta\omega)=\sum_m R_\text{I}(m,\Delta\omega)$ for (a) and (b), respectively.
Other parameters are $l=5$, and $\Omega/E_{l}=0.2$.}
\label{fig:fig6}
\end{figure}

The numerically evaluated spectra are shown in Fig.~\ref{fig:fig4}. The finite total angular momentum can be identified from the angular-momentum resolved spectroscopy [Fig.~\ref{fig:fig4}(b)]. The discretized peaks in the integrated spectra [Fig.~\ref{fig:fig4}(c)(d)] are a direct consequence of the quantized angular momentum.

\section{Molecular state for scenario II}

The molecular state of the second scenario can be written as
\begin{align}
\left|\Psi_{q}^{(2)}\right\rangle =\sum_{m}\sum_{\beta=\pm}C_{m,\beta}b_{q-m}^{\dagger}a_{m,\beta}^{\dagger}|0\rangle,
\end{align}
where $C_{m,\beta}$ is the wavefunction, and $q$ represents the total angular momentum.

From the Schr\"odinger's equation
$H^{(2)}|\Psi_q^{(2)}\rangle=E_q^{(2)}|\Psi_q^{(2)}\rangle$, the closed equation for the molecular energy is
\begin{align}
\frac{1}{g}=\underset{m}{\sum}\left(\frac{u_{m}^{2}}{E_{q}^{(2)}-\epsilon_{m-q}^b-\lambda_{m,+}}+\frac{v_{m}^{2}}{E_{q}^{(2)}-\epsilon_{m-q}^b-\lambda_{m,-}}\right),
\end{align}
where $u_m$ and $v_m$ are the same as Eqs.~(\ref{eq:u})(\ref{eq:v}).
The molecular binding energy is $E_M^{(2)}=E_{q}^{(2)}-\left[\min(\epsilon_m^b)+\min(\lambda_{m\text{,+}})\right]$.

We show the molecular energies as functions of the total angular momentum in Fig.~\ref{fig:fig5}(a)(b)(c). The ground-state molecules all have a finite angular momentum, due to the interplay of the SOAMC and spin-selective interaction. Their angular momenta are also reflected in the wavefunctions, which are asymmetric in the angular momentum space, as shown in Fig.~\ref{fig:fig5}(d)(e)(f).

As a convenient scheme to resolve the finite angular momentum, we propose to perform an inverse Raman spectroscopy with Laguerre-Gaussian beams. The Raman process couples atomic population in a spectator $|3\rangle$, to an empty state $|b\rangle$, in the presence of the SOAM coupled atoms.
The coupling Hamiltonian is written as
\begin{align}
H_{\text{R}}=\Omega_{\text{R}}\int d\mathbf{r}e^{im\theta}\psi^{\dagger}(\mathbf{r})\text{\ensuremath{\psi}}_{3}(\mathbf{r}),
\end{align}
where $m$ indicates the angular-momentum transform of the Raman probe which consists of two Laguerre-Gaussian beams with different orbital angular momentum.
Assuming atoms in state $|3\rangle$ have zero angular momentum, the Raman probe thus picks out wavefunction components $C_{q-m,\lambda}$ in the molecular wavefunction, enabling an angular-momentum resolved spectroscopy. Note that this Raman process is different from the one in the direct Raman spectroscopy for scenario I, where there is zero net angular-momentum transfer.

The angular-momentum resolved transfer rate of the inverse Raman spectroscopy is
\begin{align}
R_{\text{I}}(m,\Delta\omega)&=-\sum_{m',\beta=\pm}\text{Im}\frac{\left|\left\langle \Psi_{q}^{(2)}\left|b_{m}^{\dagger}\right|m',\beta\right\rangle \right|^{2}}{\Delta\omega-(E_{q}^{(2)}-\lambda_{m',\beta})+i0^+},\nonumber\\
&=\pi\underset{\beta=\pm}{\sum}|C_{q-m,\beta}|^{2}\delta\left[\Delta\omega+(\lambda_{q-m,\beta}-E_{q}^{(2)})\right].
\end{align}
Here $\Delta\omega$ is the two-photon detuning of the Raman probe, under the rotating frame where $|b\rangle$ and $|3\rangle$ are degenerate in the absence of SOAMC and interaction.

As illustrated in Fig.~\ref{fig:fig6}(b), the asymmetry of the inverse Raman spectrum (with respect to $m=0$) clearly reveals the finite total angular momentum of the corresponding molecular state.

\section{Summary}

We show that the unique pairing mechanism under SOAMC is reflected in the molecular state under the corresponding configurations. Using two different scenarios as examples, we demonstrate the interplay of SOAMC, Zeeman fields, and interaction can give rise to molecular states with finite angular momentum, which are the few-body angular analogue of the Spin-orbit coupling (SOC)-induced Fulde-Ferrell states in Fermi superfluids. We propose to detect the molecular angular momentum through direct and inverse Raman spectroscopy. Since molecular states persist at temperatures higher than the superfluid temperature, they are an ideal candidate for experimental detection, particularly in light of heating introduced by SOAMC.
Further, in contrast to the SOC-induced finite center-of-mass momentum pairing states (either pairing superfluid or molecules)~\cite{socreview4}, the angular momentum of the molecular states under study here are quantized, and are therefore more accessible to experimental detection. Therefore, our work provides a practical route toward the confirmation of the unique pairing mechanism under synthetic gauge fields such as SOC and SOAMC.

\section*{Acknowledgements}
This work has been supported by the National Natural Science Foundation of China (11974331, 12104406, 11974384) and the National Key R\&D Program (Grant Nos. 2017YFA0304100). K.C. acknowledges support from the
startup grant of Zhejiang Sci-Tech University (Grant No. 21062338-Y). S.P. acknowledges support from the Natural Science Foundation of Hubei Province (Grant No 2021CFA027).

\bibliographystyle{apsrev4-1}

\begin{thebibliography}{99}
\bibitem{Lin-11} Y.-J. Lin, K. Jim\'{e}nez-Garc\'{i}a, and I. B. Spielman,
Nature (London) {\bf 471}, 83 (2011).

\bibitem{Zhang-12}
P. Wang, Z.-Q. Yu, Z. Fu, J. Miao, L. Huang, S. Chai, H. Zhai, and J. Zhang,
Phys. Rev. Lett. {\bf 109}, 095301 (2012).

\bibitem{Zwierlein-12}
L. W. Cheuk, A. T. Sommer, Z. Hadzibabic, T. Yefsah, W. S. Bakr, and M. W. Zwierlein,
Phys. Rev. Lett. {\bf 109}, 095302 (2012).

\bibitem{socreview1} V. Galitski and I. B. Spielman, Nature (London) {\bf 494}, 49 (2013).

\bibitem{socreview2} N. Goldman, G. Juzeli\={u}nas, and P. \"Ohberg, and I. B. Spielman, Rep. Prog. Phys. {\bf 77}, 126401 (2014).

\bibitem{socreview3} H. Zhai, Rep. Prog. Phys. {\bf78}, 026001 (2015).

\bibitem{socreview4} W. Yi, W. Zhang, and X. Cui, Sci. China Phys. Mech. Astron. {\bf 58}, 1 (2015).

\bibitem{socreview5} J. Zhang, H. Hu, X. J. Liu, and H. Pu, Annu. Rev. Cold At. Mol. {\bf 2}, 81 (2014).

\bibitem{socreview6} L. Zhang and X. J. Liu, in \textit{Synthetic Spin-Orbit Coupling in Cold Atoms}, edited by W. Zhang, W. Yi, and C. A.R. S\'a  Melo (World Scientific, Singapore, 2018), pp.1--87.

\bibitem{Jiang-19}
D. Zhang, T. Gao, P. Zou, L. Kong, R. Li, X. Shen, X.-L. Chen, S.-G. Peng, M. Zhan, H. Pu, and K. Jiang,
Phys. Rev. Lett. {\bf 122}, 110402 (2019).

\bibitem{Lin-18}
H.-R. Chen, K.-Y. Lin, P.-K. Chen, N.-C. Chiu, J.-B. Wang, C.-A. Chen, P.-P. Huang, S.-K. Yip, Y. Kawaguchi, and Y.-J. Lin,
Phys. Rev. Lett. {\bf 121}, 113204 (2018).

\bibitem{Hu-15}
Y.-X. Hu, C. Miniatura, and B. Gr\'{e}maud,
Phys. Rev. A {\bf 92}, 033615 (2015)

\bibitem{Pu-15}
M. DeMarco and H. Pu,
Phys. Rev. A  {\bf 91}, 033630 (2015).

\bibitem{Sun-15}
K. Sun, C. Qu, and C. Zhang,
Phys. Rev. A {\bf 91}, 063627 (2015).

\bibitem{Qu-15}
C. Qu, K. Sun, and C. Zhang,
Phys. Rev. A {\bf 91}, 053630 (2015).

\bibitem{Chen-16}
 L. Chen, H. Pu, and Y. Zhang,
 Phys. Rev. A {\bf 93}, 013629 (2016).

\bibitem{Hu-19}
 X.-L. Chen, S.-G. Peng, P. Zou, X.-J. Liu, and H. Hu, Phys. Rev. Research {\bf 2}, 033152 (2020).

\bibitem{Chen-19}
K.-J. Chen, F. Wu, J. Hu, and L. He, Phys. Rev. A {\bf 102}, 013316 (2020).

\bibitem{Han-20}
L. Chen, Y. Zhang, and H. Pu, Phys. Rev. Lett. {\bf 125}, 195303 (2020).

\bibitem{Duan-20}
Y. Duan, Y. M. Bidasyuk, and A. Surzhykov, Phys. Rev. A {\bf 102}, 063328 (2020).

\bibitem{Chen-20}
K.-J. Chen, F. Wu, S.-G. Peng, W. Yi, and L. He, Phys. Rev.
Lett. {\bf 125}, 260407 (2020).

\bibitem{Wang-21}
 L.-L. Wang, A.-C. Ji, Q. Sun, and J. Li,
 Phys. Rev. Lett. {\bf 126}, 193401 (2021).

\bibitem{Chen-22} K.-J. Chen, F. Wu, L. He, and W. Yi, arXiv:2201.12486.

\bibitem{heating} X. Cui, B. Lian, T.-L. Ho, B. L. Lev, and H. Zhai, Phys. Rev. A 88, 011601(R) (2013).

\bibitem{spielmandressed} R. A. Williams, M. C. Beeler, L. J. LeBlanc, K. Jimenez-Garcia, and I. B. Spielman, Phys. Rev. Lett. 111, 095301 (2013).

\bibitem{zhangdressed} Z. Fu, L. Huang, Z. Meng, P. Wang, L. Zhang, S. Zhang, H. Zhai, P. Zhang, and J. Zhang, Nat. Phys. 10, 110 (2014).

\bibitem{Wu-13} F. Wu, G.-C. Guo, W. Zhang, and W. Yi, Phys. Rev. Lett. {\bf 110}, 110401 (2013).

\bibitem{ZCY-14} L. Zhou, X. Cui, and W. Yi, Phys. Rev. Lett. 112, 195301 (2014).

\bibitem{Peng-12} S.-G. Peng, X.-J. Liu, H. Hu, and K. Jiang, Phys. Rev. A 86, 063610 (2012).
\end{thebibliography}

\end{document}